\documentclass[prb,twocolumn,groupedaddress]{revtex4}
\usepackage{amssymb}
\usepackage{amsmath}
\usepackage{epsfig}
\usepackage{eufrak}
\usepackage{graphicx}
\usepackage{verbatim}

\begin{document}



\title{Quasi-planar optics: computing light propagation and scattering in planar waveguide arrays}

\author{Sukosin Thongrattanasiri}
\author{Justin Elser}
\author{Viktor A. Podolskiy}

\email{vpodolsk@physics.oregonstate.edu}

\address{
Department of Physics, 
301 Weniger Hall, Oregon State University, 
Corvallis OR 97331}

\begin{abstract}
We analyze wave propagation in coupled planar waveguides, pointing specific attention to modal cross-talk and out-of-plane scattering in quasi-planar photonics. An algorithm capable of accurate numerical computation of wave coupling in arrays of planar structures is developed and illustrated on several examples of plasmonic and volumetric waveguides. An analytical approach to reduce or completely eliminate scattering and modal cross-talk in planar waveguides with anisotropic materials is also presented. 
\end{abstract}


\maketitle

\section{Introduction}
On-chip communications, surface plasmon optics, and Si photonics are all examples of {\it planar optics}, where the optical radiation is controllably guided on the plane of a photonic chip. A number of planar optical elements including lenses, mirrors, and on-chip waveguides, both plasmonic and dielectric have been recently designed, fabricated, and characterized\cite{bozhevolnyiprl,maiernatmat,smolyaninovprl,shinfan,bozhevolnyinature,barnesnature,stockmanprl,zianaturenano,liuapl,smolyaninovprb,vlasov,boardmanBook}. However, with a few exceptions\cite{smolyaninovprb,oultonPRB,schevchenkoBook}, the majority of recent studies focuses on in-plane propagation of light and neglects out-of-plane scattering of radiation.
In this work we analyze the out-of-plane light scattering and modal cross-talk due to effective index change inside planar waveguides and demonstrate that this scattering may substantially affect the propagation of confined modes in complex planar systems. We design a numerical approach to solve the problem of scattering and modal cross-talk in planar or quasi-planar structures that contain anisotropic elements, and present the technique to reduce or completely eliminate the scattering and cross-talk with anisotropic metamaterials\cite{elserprl}. 

The problem of out-of-plane scattering has been of consistent interest to photonic community. Although a number of finite-difference and finite-elements techniques, available today, can successfully solve the problem of scattering in relatively small geometries ($\lesssim 10\lambda_0$, with $\lambda_0$ being free-space wavelength), analysis of wave propagation in an extended system is beyond the capabilities of methods that rely on finite-size meshing of space/time. One of the ways to reduce memory requirements to calculate the field in an extended structure with moderate number of scattering interfaces is to implement some sort of wave-matching technique where the modal spectrum is constructed to satisfy the solutions of Maxwell's equations in the space, and only boundary conditions {\it at scattering interfaces} are enforced, resulting in calculations of {\it amplitudes} of the modes. Effectively, modal expansion can replace the need to calculate all field components at every {\it point of space} with the need to calculate modal amplitudes in every {\it region of space}.

\begin{figure*}
\centerline{\includegraphics[width=13cm]{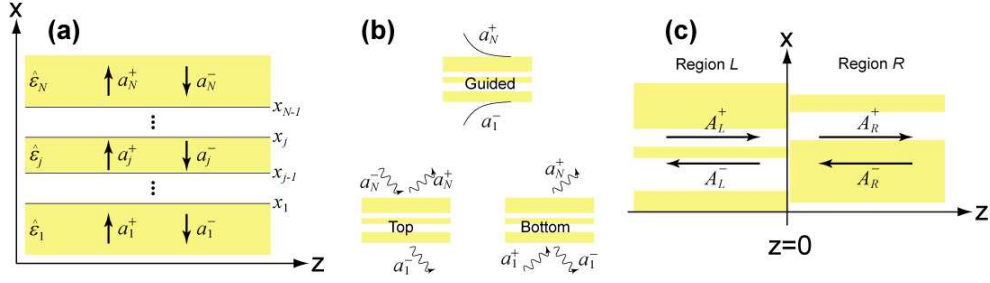}}
\vspace{10pt}
\caption{\label{figGeom} (color online) Schematic geometry of the multi-layered structures and electromagnetic mode types used in the manuscript. Geometry of the single multilayer stack is shown in (a); panel (b) explains the composition of top, bottom, and guided modes; field profiles in the outside layers of the multilayer are shown; the interface between two multilayer stacks is shown in (c) }
\end{figure*}

One of the first descriptions of wave-matching approach and its applications for highly-conductive plasmonic guides can be found in Refs.\cite{clarricoats,schevchenkoBook}. Ref.\cite{schevchenkoBook} also describes scattering by planar guides with highly symmetric crossections. An approach to calculate to modal cross-talk and scattering in 1D guides was developed in Refs.\cite{rozziBook}. Scattering by periodically corrugated systems has been analyzed with rigorous coupled wave analysis (RCWA) in\cite{moharam}. Recently, the generalization of field expansion to calculate scattering in plasmonic planar guides has been presented in\cite{oultonPRB}. However, while the mode-matching calculations were proven to be highly efficient, the technique had failed in the proximity to plasmon resonance condition, when the field of a surface wave is highly confined to the proximity of metal interface. Here we present a wave-matching technique that is capable of solving for wave scattering in complex systems formed by coupled planar waveguides. 

The rest of the manuscript is organized as follows. In Section 2, we present the mode structure of an arbitrary planar guide used in this work (Fig.\ref{figGeom}). Section 3 is devoted to the development of mode-matching technique in quasi-planar system comprising a uniform in $y$-direction array of planar guides. The presented numerical approach is illustrated on examples of light propagation in several plasmonic and metamaterial systems in Section 4. Finally, Section 5 develops the formalism of truly planar photonics where out-of-plane scattering and modal cross-talk are not possible and presents an approach to design extremely-low-scattering plasmonic circuits.

In this work we use the following notations for the electric and magnetic fields in the system: the total electric $(E)$ and magnetic $(H)$ fields are shown in italic font; the fields of modal components contributing to the total field are represented with symbols $\mathbb{E}$ and $\mathbb{H}$; the components of an individual mode in a particular layer of the multilayer stack are represented with calligraphic symbols $\mathcal{E}$ and $\mathcal{H}$. 

\section{Modal spectrum of planar guides} 

We start from analysis of modal spectrum of a planar waveguide, schematically shown in Fig.\ref{figGeom}(a). The structure comprises a set of $N$ planar layers with layer interfaces parallel to $yz$ plane, with $j^{\rm th}$ layer occupying the space between $x_{j-1}<x\le x_j$, and having (uniaxial) dielectric permittivity described by a diagonal tensor with diagonal components $\hat{\epsilon_j}=\{\epsilon^{xx}_j,\epsilon^{yz}_j,\epsilon^{yz}_j\}$. We assume that $x_0=-\infty$, and $x_N=\infty$. In this section we consider the structure that is infinitely extended in the $yz$ plane.

The electromagnetic fields in this layered system can be represented as a set of transverse-electric (TE) and transverse-magnetic (TM) polarized waves (modes). Each mode of the multilayer constitutes a solution of Maxwell's equations that is finite for $-\infty<x<\infty$\cite{bornwolf}. In homogeneous layered structure the mode can be parameterized by a combination of (i) its polarization (TE/TM), (ii) the in-plane components of the wavevector ($k_y,k_z$), and (iii) a set of layer-specific complex coefficients $\{a_j^\pm\}$ playing the role of amplitudes of the mode components: 
\begin{widetext}
\begin{eqnarray}
\vec{\mathbb{E}}(k_y,k_z)=\left\{
\begin{array}{ll}
a_1^+ \vec{\mathcal{E}}_1(k_{x;1} (k_y,k_z), k_y, k_z)+ a_1^- \vec{\mathcal{E}}_1(-k_{x;1}(k_y,k_z),k_y,k_z),&x<x_1
\\ \ldots
\\
a_j^+ \vec{\mathcal{E}}_j(k_{x;j} (k_y,k_z),k_y,k_z)+ a_j^- \vec{\mathcal{E}}_j(-k_{x;j} (k_y,k_z) ,k_y,k_z),&x_{j-1}<x<x_j
\\ \ldots
\\
a_N^+ \vec{\mathcal{E}}_N(k_{x;N} (k_y,k_z),k_y,k_z)+ a_N^- \vec{\mathcal{E}}_N(-k_{x;N} (k_y,k_z),k_y,k_z),&x_{N-1}<x
\end{array}
\right.
\\
\vec{\mathbb{H}}(k_y,k_z)=\left\{
\begin{array}{ll}
a_1^+ \vec{\mathcal{H}}_1(k_{x;1} (k_y,k_z),k_y,k_z)+ a_1^- \vec{\mathcal{H}}_1(-k_{x;1}(k_y,k_z),k_y,k_z),&x<x_1
\\ \ldots
\\
a_j^+ \vec{\mathcal{H}}_j(k_{x;j} (k_y,k_z),k_y,k_z)+ a_j^- \vec{\mathcal{H}}_j(-k_{x;j} (k_y,k_z),k_y,k_z),&x_{j-1}<x<x_j
\\ \ldots
\\
a_N^+ \vec{\mathcal{H}}_N(k_{x;N} (k_y,k_z),k_y,k_z)+ a_N^- \vec{\mathcal{H}}_N(-k_{x;N} (k_y,k_z),k_y,k_z),&x_{N-1}<x
\end{array}
\right.
\end{eqnarray}
where the fields and dispersion relations in each layer are described by: 
\begin{eqnarray}
{\rm TE-polarized\; waves:}&\left\{
\begin{array}{l}
\begin{array}{lllcr}
\vec{\mathcal{E}}_j&=\frac{ e^{-i\omega t+i \vec{k}\cdot \vec{r}}}{\sqrt{k_y^2+k_z^2}} &\{0, &k_z, &-k_y\}
\\
\vec{\mathcal{H}}_j&=\frac{ e^{-i\omega t+i \vec{k}\cdot \vec{r}}c}{\omega\sqrt{k_y^2+k_z^2}}&\{-(k_y^2+k_z^2),&k_{x;j} k_y, &k_{x;j} k_z \}
\end{array}
\\ \frac{k_{x;j}^2+k_y^2+k_z^2}{\epsilon^{yz}_j}= \frac{\omega^2}{c^2}
\end{array}
\right.
\\
{\rm TM-polarized\; waves:}
&\left\{
\begin{array}{l}
\begin{array}{lllcr}
\vec{\mathcal{E}}_j &= \frac{ e^{-i\omega t+i \vec{k}\cdot \vec{r}}}{k_y^2+k_z^2} &\{k_y^2+k_z^2, &-\frac{\epsilon_j^{xx} k_{x;j} k_y }{\epsilon_j^{yz}},& - \frac{\epsilon^{xx}_j k_{x;j} k_z }{\epsilon_j^{yz}}\}
\\
\vec{\mathcal{H}}_j &= \frac{ e^{-i\omega t+i \vec{k}\cdot \vec{r}} \omega}{(k_y^2+k_z^2)c}&\{0, &\epsilon_j^{xx}k_z, &-\epsilon_j^{xx}k_y\}
\end{array}
\\ \frac{k_{x;j}^2}{\epsilon^{yz}_j}+\frac{k_y^2+k_z^2}{\epsilon^{xx}_j}=\frac{\omega^2}{c^2}
\end{array}
\right.
\end{eqnarray}
\end{widetext}
Note that for a given mode only two of the amplitudes $a_j^\pm$ are independent of each other. Indeed, the remaining amplitudes can be calculated using the well-known {\it transfer matrix} method\cite{bornwolf}:
\begin{eqnarray}
\left(
\begin{array}{c}
a^-_{j+1}
\\
a^+_{j+1}
\end{array}
\right)
=\alpha_{j}\left[
\begin{array}{cc}
(1+K_{j}) \phi_j^- & (1-K_{j}) \phi_j^+
\\
(1-K_{j}) /\phi_j^+ & (1+K_{j}) /\phi_j^-
\end{array}
\right]
\left(
\begin{array}{c}
a^-_{j}
\\
a^+_{j}
\end{array}
\right),
\end{eqnarray} 
with polarization-dependent parameters $\phi_j,\alpha_j$, and $K_j$ given by $\phi_j^\pm=\exp[i (k_{x;j+1}\pm k_{x;j})x_j]$, $\alpha_j^{\rm TE}=1/2,\; \alpha_j^{\rm TM}=\epsilon_j^{xx}/(2\epsilon_{j+1}^{xx}),\; K_j^{\rm TE}=k_{x;j}/k_{x;j+1},\; K_j^{\rm TM}=k_{x;j}\epsilon^{yz}_{j+1}/(k_{x;j+1}\epsilon^{yz}_j)$.

For multilayer systems, layer-specific transfer matrices can be multiplied together, yielding the transfer-matrix relating the fields in any two layers of the multilayer stack, and thus solving the problem of reflection and transmission of a plane wave by the multilayer composite. The singular solution that corresponds non-zero scattered waves on both sides of the multilayer structure ($a^-_1, a^+_N\neq 0$) with zero incident fields ($a^+_1=a^-_N= 0$) corresponds to eigen (guided) mode of the stack\cite{bornwolf}.

For each polarization, full spectrum of modes supported by the stack includes three groups of waves [see Fig.\ref{figGeom}(b)]. The first group contains a discrete spectrum of guided modes, exponentially decaying into first and last layers ($a_1^+=a_N^-=0,\;a_1^-=1$). Here we characterize these waves by the in-plane components of their wavevectors $\{k_y,k_z\}$. 

The remaining groups of modes contain the continuum of waves, known as open-waveguide modes\cite{schevchenkoBook,oultonPRB} (bulk modes). The first of these groups represents the modes originated by a plane wave that is incident on the layered structure from the top layer ($a_1^+=0,\;a_N^-=1$), while the second group represents the wave incident on the structure from the bottom layer  ($a_1^+=1,\;a_N^-=0$). Here, the modes of the first group (``top modes'') are parameterized by the real-valued $k_{x;N}$ in the top ($N^{th}$) layer, while the ``bottom modes'' are parameterized by the real-valued $k_{x;1}$ in the bottom ($1^{st}$) layer.

In the limit of symmetric distribution of permittivity $\hat{\epsilon}(x)=\hat{\epsilon}(-x)$, the proposed here spectrum of top and bottom modes is equivalent to the proposed earlier\cite{schevchenkoBook} combinations of ``standing wave'' modes with symmetric and anti-symmetric $x$-profiles. However, in contrast to the latter, the combination of top and bottom modes is more easily generalizable to the case of non-symmetric (such as plasmonic) planar guides. Note that in majority of previous studies of plasmonic structures\cite{oultonPRB}, bottom modes were explicitly omitted. As explained below, this omission becomes crucial in the regime of strong surface plasmon polariton (SPP) scattering, e.g. in proximity to SPP resonance or in plasmonic step geometry (Fig.\ref{figSPPscat}).

Overall, the field inside the guiding structure can be written as: 
\begin{widetext}
\begin{eqnarray}
\label{eqModes}
\begin{array}{lrl}
\vec{E}=&\sum_q &\left[ A^{{(q)}^+} \vec{\mathbb{E}}(k_y,k_z^{(q)})+A^{{(q)}^-} \vec{\mathbb{E}}(k_y,-k_z^{(q)})\right]
\\ &+\int_{0}^\infty &\left[A^{\rm top^+}(k_x)\vec{\mathbb{E}}(k_y,k_z(k_{x;N}))
+A^{\rm top^-}(k_x)\vec{\mathbb{E}}(k_y,-k_z(k_{x;N}))
\right.
\\&&+A^{\rm btm^+}(k_x)\vec{\mathbb{E}}(k_y,k_z(k_{x;1}))
\left.+A^{\rm btm^-}(k_x)\vec{\mathbb{E}}(k_y,-k_z(k_{x;1}))
\right] dk_x 
\end{array}
\\
\nonumber
\begin{array}{lrl}
\vec{H}=&\sum_q &\left[ A^{{(q)}^+} \vec{\mathbb{H}}(k_y,k_z^{(q)})+A^{{(q)}^-} \vec{\mathbb{H}}(k_y,-k_z^{(q)})\right]
\\ &+\int_{0}^\infty &\left[A^{\rm top^+}(k_x)\vec{\mathbb{H}}(k_y,k_z(k_{x;N}))
+A^{\rm top^-}(k_x)\vec{\mathbb{H}}(k_y,-k_z(k_{x;N}))
\right.
\\&&+A^{\rm btm^+}(k_x)\vec{\mathbb{H}}(k_y,k_z(k_{x;1}))
\left.+A^{\rm btm^-}(k_x)\vec{\mathbb{H}}(k_y,-k_z(k_{x;1}))
\right] dk_x. 
\end{array}
\end{eqnarray}
\end{widetext}

Here we assume that all the modes in the layered material have the same value of $k_y$. This assumption does not limit the applicability of the developed technique since due to translational symmetry, any solution of Maxwell's equations in the coupled waveguide set can be represented as a linear combination of solutions for specific values of $k_y$. Likewise, we assume that excitation and response of the system are monochromatic $[E,H\propto \exp(-i\omega t)]$. The linearity of Maxwell's equations makes it possible to generalize the developed formalism for the arbitrary pulse excitation by representing the incident radiation by linear combination of monochromatic waves. 

Note that in the process of calculating the waveguide modes, it may be necessary to determine the proper sign of the $k_x$ (or $k_z$) component of the wavevector in a particular layer. If the component of the wavevector has complex value, this sign is determined from the requirement for the mode to be finite in its domain. If the wavevector component is real, its sign should be determined to impose the propagation of energy in the positive $x$ (or $z$) direction\cite{govyadinovAPL}. 

The set of waveguide modes defined above allows the introduction of the scalar product:
\begin{eqnarray}
<\vec{\mathbb{E}}_1|\vec{\mathbb{H}}_2^\dagger>=\int_{-\infty}^\infty (\vec{\mathbb{E}}_1\times \vec{\mathbb{H}}_2^\dagger)\cdot\hat{z}dx
\end{eqnarray}
where dagger ($\dagger$) corresponds to the adjoined field - field of the mode propagating in the reversed $z$ direction\cite{rozziBook,schevchenkoBook}. It can be shown\cite{schevchenkoBook,rozziBook,oultonPRB} that in a given multilayer (i) all TM-polarized waves are orthogonal to all TE-polarized waves, (ii) the guided modes are orthogonal to each other, and (iii) the top and bottom modes may have some coupling, depending on $\hat{\epsilon}_1$ and $\hat{\epsilon}_N$: if $\hat{\epsilon}_1=\hat{\epsilon}_N$, the top and bottom modes corresponding to the same value of $k_x$ are coupled to each other and are orthogonal to all other modes; if one of the two materials is lossy (as it is usually the case with plasmonic structures), the top and bottom modes are, as a rule, orthogonal to each other. 

Note that similar to what has been suggested in Ref.\cite{oultonPRB} the scalar product can be calculated analytically, significantly speeding up the calculation.

\section{Mode coupling across multilayer stacks }
\subsection{General formalism} 
We now turn to the main point of this work -- discussion of coupling of the modes at the boundary between the two multilayer structures. For simplicity, we present results for the case when the interface is located at $z=0$ [Fig.\ref{figGeom}(c)]. Generalization of the technique for other locations of the interface is straightforward. 

We are solving the classical scattering problem: finding the fields scattered by the interface provided that the incident fields are known. The incident fields are represented by the modes propagating in the $+z$ direction on the left-hand-side of the interface ($z<0$) and by the modes travelling in the $-z$ direction on the right-hand-side of the interface ($z>0$); the scattered fields are represented by the modes travelling in the $-z$ direction on the left-hand-side of the interface and by the modes travelling in the $+z$ direction on the right-hand-side of the interface. The modal representation [Eqs.(\ref{eqModes})] reduces the scattering problem to an arithmetic task of finding the coefficients $A_L^-$ and $A_R^+$ as a function of $A_L^+$ and $A_R^-$, which can be solved by imposing the following set of boundary conditions: 
\begin{eqnarray}
\label{eqBCond}
E_{L_x}=E_{R_x}; 
& H_{L_y}=H_{R_y},
\\
\nonumber
E_{L_y}=E_{R_y}; 
& H_{L_x}=H_{R_x}.
\end{eqnarray}
As can be explicitly verified, the remaining boundary conditions follow from Eqs.(\ref{eqBCond}). 

In the case of normal incidence ($k_y=0$), TM- and TE- modes do not couple to each other. Correspondingly, in this case the first two boundary conditions in Eqs.(\ref{eqBCond}) describe the reflection, transmission, and scattering of TM-polarized waves, while the remaining two conditions describe the optical properties of TE-polarized modes. 

In order to solve the scattering problem, Eqs.(\ref{eqBCond}) need to be converted into the set of coupled linear equations for the amplitudes of the scattered modes. To achieve this goal, we substitute the modal expansion [Eqs.(\ref{eqModes})] in Eqs.(\ref{eqBCond}), and subsequently multiply the resulting expressions by the adjoined fields of left- and right-hand-side modes, as illustrated below. 

\subsection{Numerical implementation of the algorithm} 

In numerical simulations, it is necessary to replace the continuous integration over $k_x$ with finite sums. Thus, Eqs.(\ref{eqModes}) becomes
\begin{eqnarray}
\label{eqModesDiscrete}
\vec{E}=\sum_m [A^{(m)^+} \vec{\mathbb{E}}^{(m)^+}+ A^{(m)^-} \vec{\mathbb{E}}^{(m)^-}] w^{(m)}
\\
\nonumber
\vec{H}=\sum_m [A^{(m)^+} \vec{\mathbb{H}}^{(m)^+}+ A^{(m)^-} \vec{\mathbb{H}}^{(m)^-}] w^{(m)} 
\end{eqnarray}
where $\vec{\mathbb{E}}^{(m)^\pm}\equiv \vec{\mathbb{E}}(k_y,\pm k_z^{(m)})$, and similar for $\vec{\mathbb{H}}$; the summation in Eqs.(\ref{eqModesDiscrete}) goes over all modes (guided, top, and bottom), and the weight factors $w$ are equal to $1$ for the guided modes, and are determined by the integration method used for top and bottom modes\cite{cpBook}. Note that the number of modes on the left-hand-side of the interface does not necessarily equal to the number of modes on the right-hand-side of the interface. 

Eqs.(\ref{eqBCond}) now become: 
\begin{widetext}
\begin{eqnarray}
\nonumber
\sum_{m=1}^{N_L} [A^{(m)^+}_L \mathbb{E}_{L_x}^{(m)^+}+ A^{(m)^-}_L  {\mathbb{E}}_{L_x}^{(m)^-}] w^{(m)}_L
=\sum_{m=1}^{N_R}  [A^{(m)^+}_R \mathbb{E}_{R_x}^{(m)^+}+ A^{(m)^-}_R {\mathbb{E}}_{R_x}^{(m)^-}] w^{(m)}_R, 
\\
\nonumber
\sum_{m=1}^{N_L}  [A^{(m)^+}_L \mathbb{H}_{L_y}^{(m)^+}+ A^{(m)^-}_L  {\mathbb{H}}_{L_y}^{(m)^-}] w^{(m)}_L
=\sum_{m=1}^{N_R} [A^{(m)^+}_R \mathbb{H}_{R_y}^{(m)^+}+ A^{(m)^-}_R {\mathbb{H}}_{R_y}^{(m)^-}] w^{(m)}_R
\\
\label{eqBCondDiscrete}
\sum_{m=1}^{N_L}  [A^{(m)^+}_L \mathbb{E}_{L_y}^{(m)^+}+ A^{(m)^-}_L  {\mathbb{E}}_{L_y}^{(m)^-}] w^{(m)}_L
=\sum_{m=1}^{N_R} [A^{(m)^+}_R \mathbb{E}_{R_y}^{(m)^+}+ A^{(m)^-}_R {\mathbb{E}}_{R_y}^{(m)^-}] w^{(m)}_R
\\
\nonumber
\sum_{m=1}^{N_L}  [A^{(m)^+}_L \mathbb{H}_{L_x}^{(m)^+}+ A^{(m)^-}_L  {\mathbb{H}}_{L_x}^{(m)^-}] w^{(m)}_L
=\sum_{m=1}^{N_R} [A^{(m)^+}_R \mathbb{H}_{R_x}^{(m)^+}+ A^{(m)^-}_R {\mathbb{H}}_{R_x}^{(m)^-}] w^{(m)}_R
\end{eqnarray}
To solve for $N_R+N_L$ unknown amplitudes, we multiply first two equations in Eqs.(\ref{eqBCondDiscrete}) by the fields of TM-polarized modes and integrate the resulting products over $x$; similarly, we multiply the two latter equations by the fields of TE-polarized modes and perform the integration. Assuming that the index $m$ first spans the TE-polarized and then TM-polarized waves, the procedure results in the following two sets of matrix equations:
\begin{eqnarray}
\label{eqRMats}
\widehat{\mathfrak{E}_{\mathfrak{R}L}^+}^{mn} A_L^{(m)^+}+ \widehat{\mathfrak{E}_{\mathfrak{R}L}^-}^{mn} A_L^{(m)^-}
=\widehat{\mathfrak{E}_{\mathfrak{R}R}^+}^{mn} A_R^{(m)^+}
+\widehat{\mathfrak{E}_{\mathfrak{R}R}^-}^{mn} A_R^{(m)^-}
\\
\nonumber
\widehat{\mathfrak{H}_{\mathfrak{R}L}^+}^{mn} A_L^{(m)^+}+ \widehat{\mathfrak{H}_{\mathfrak{R}L}^-}^{mn} A_L^{(m)^-}
=\widehat{\mathfrak{H}_{\mathfrak{R}R}^+}^{mn} A_R^{(m)^+}
+\widehat{\mathfrak{H}_{\mathfrak{R}R}^-}^{mn} A_R^{(m)^-}
\end{eqnarray}
and 
\begin{eqnarray}
\label{eqLMats}
\widehat{\mathfrak{E}_{\mathfrak{L}L}^+}^{mn} A_L^{(m)^+}+ \widehat{\mathfrak{E}_{\mathfrak{L}L}^-}^{mn} A_L^{(m)^-}
=\widehat{\mathfrak{E}_{\mathfrak{L}R}^+}^{mn} A_R^{(m)^+}
+\widehat{\mathfrak{E}_{\mathfrak{L}R}^-}^{mn} A_R^{(m)^-}
\\
\nonumber
\widehat{\mathfrak{H}_{\mathfrak{L}L}^+}^{mn} A_L^{(m)^+}+ \widehat{\mathfrak{H}_{\mathfrak{L}L}^-}^{mn} A_L^{(m)^-}
=\widehat{\mathfrak{H}_{\mathfrak{L}R}^+}^{mn} A_R^{(m)^+}
+\widehat{\mathfrak{H}_{\mathfrak{L}R}^-}^{mn} A_R^{(m)^-}
\end{eqnarray}
where the summation over repeated index $m$ is assumed and matrix elements are given by: 
\begin{eqnarray}
\nonumber
\mathfrak{E}_{\mathfrak{R}\{L|R\}}^{\pm^{mn}}
&=w^{(m)}_{\{L|R\}}\left\{
\begin{array}{ll}
\int_{-\infty}^\infty \mathbb{E}_{\{L|R\} _y}^{(m)^\pm} \mathbb{H}_{R_x}^{(n)^-} dx, & n\le N_R^{\rm TE}
\\
\int_{-\infty}^\infty \mathbb{E}_{\{L|R\}_x}^{(m)^\pm} \mathbb{H}_{R_y}^{(n)^-}dx, & n>N_R^{\rm TE}
\end{array}
\right., 
\\
\mathfrak{H}_{\mathfrak{R}\{L|R\}}^{\pm^{mn}}
&=w^{(m)}_{\{L|R\}}\left\{
\begin{array}{ll}
\int_{-\infty}^\infty \mathbb{H}_{\{L|R\} _x}^{(m)^\pm} \mathbb{E}_{R_y}^{(n)^-}dx, & n\le N_R^{\rm TE}
\\
\int_{-\infty}^\infty \mathbb{H}_{\{L|R\}_y}^{(m)^\pm} \mathbb{E}_{R_x}^{(n)^-}dx, & n>N_R^{\rm TE}
\end{array}
\right., 
\\
\nonumber
\mathfrak{E}_{\mathfrak{L}\{L|R\}}^{\pm^{mn}}
&=w^{(m)}_{\{L|R\}}\left\{
\begin{array}{ll}
\int_{-\infty}^\infty \mathbb{E}_{\{L|R\} _y}^{(m)^\pm} \mathbb{H}_{L_x}^{(n)^-} dx, & n\le N_R^{\rm TE}
\\
\int_{-\infty}^\infty \mathbb{E}_{\{L|R\}_x}^{(m)^\pm} \mathbb{H}_{L_y}^{(n)^-}dx, & n>N_R^{\rm TE}
\end{array}
\right., 
\\
\nonumber
\mathfrak{H}_{\mathfrak{L}\{L|R\}}^{\pm^{mn}}
&=w^{(m)}_{\{L|R\}}\left\{
\begin{array}{ll}
\int_{-\infty}^\infty \mathbb{H}_{\{L|R\} _x}^{(m)^\pm} \mathbb{E}_{L_y}^{(n)^-}dx, & n\le N_R^{\rm TE}
\\
\int_{-\infty}^\infty \mathbb{H}_{\{L|R\}_y}^{(m)^\pm} \mathbb{E}_{L_x}^{(n)^-}dx, & n>N_R^{\rm TE}
\end{array}
\right.
\end{eqnarray} 
Note that the modes of $z<0$ region are not necessarily orthogonal to the modes in the $z>0$ region. Thus, the matrices $\widehat{\mathfrak{E}^\pm_{\mathfrak{R}L}},\widehat{\mathfrak{H}^\pm_{\mathfrak{R}L}},\widehat{\mathfrak{E}^\pm_{\mathfrak{L}R}}$, $\widehat{\mathfrak{H}^\pm_{\mathfrak{L}R}}$  may have substantial non-diagonal components describing cross-talk of the modes across the interface.  

In fact, the above matrices are square and invertible only when $N_L=N_R$, in which case one of Eqs.(\ref{eqRMats},\ref{eqLMats}) can provide the information required to solve the scattering problem. However, even in this case inversion procedure may lead to significant numerical problems, and is undesirable. When $N_L\neq N_R$ these matrices are {\it rectangular} and thus, even theoretically, cannot be inverted. To overcome this difficulty, we reduce Eqs.(\ref{eqRMats},\ref{eqLMats}) to the following set of equations that represent the generalization of transfer-matrix formalism for coupled waveguide structures: 
\begin{eqnarray}
\label{eqTMMgen}
\left\{
\begin{array}{ll}
\vec{A}_R^{-}&=\widehat{\mathfrak{R}_{11}}\vec{A}_L^-+\widehat{\mathfrak{R}_{12}}\vec{A}_L^+
\\ 
\vec{A}_R^{+}&=\widehat{\mathfrak{R}_{21}}\vec{A}_L^-+\widehat{\mathfrak{R}_{22}}\vec{A}_L^+
\end{array}
\right.
\\ 
\left\{
\begin{array}{ll}
\vec{A}_L^{-}&=\widehat{\mathfrak{L}_{11}}\vec{A}_R^-+\widehat{\mathfrak{L}_{12}}\vec{A}_R^+
\\ \nonumber
\vec{A}_L^{+}&=\widehat{\mathfrak{L}_{21}}\vec{A}_R^-+\widehat{\mathfrak{L}_{22}}\vec{A}_R^+
\end{array}
\right.
\end{eqnarray}
where $\vec{A}^\pm_{\{L|R\}}\equiv\left\{A^{(1)^\pm}_{\{L|R\}},\ldots,A^{(N_{\{L|R\}})^\pm}_{\{L|R\}}\right\}$ and 
\begin{eqnarray}
&\left\{
\begin{array}{l}
\widehat{\mathfrak{R}_{11}}=
\left[\widehat{\mathfrak{E}_{\mathfrak{R}R}^+}^{-1}\widehat{\mathfrak{E}_{\mathfrak{R}R}^-}-\widehat{\mathfrak{H}_{\mathfrak{R}R}^+}^{-1}\widehat{\mathfrak{H}_{\mathfrak{R}R}^-}
\right]^{-1}
\left( \widehat{\mathfrak{E}_{\mathfrak{R}R}^+}^{-1}\widehat{\mathfrak{E}_{\mathfrak{R}L}^-}-
\widehat{\mathfrak{H}_{\mathfrak{R}R}^+}^{-1}\widehat{\mathfrak{H}_{\mathfrak{R}L}^-} \right)
\\
\widehat{\mathfrak{R}_{12}}=
\left[\widehat{\mathfrak{E}_{\mathfrak{R}R}^+}^{-1}\widehat{\mathfrak{E}_{\mathfrak{R}R}^-}-\widehat{\mathfrak{H}_{\mathfrak{R}R}^+}^{-1}\widehat{\mathfrak{H}_{\mathfrak{R}R}^-}
\right]^{-1}
\left( \widehat{\mathfrak{E}_{\mathfrak{R}R}^+}^{-1}\widehat{\mathfrak{E}_{\mathfrak{R}L}^+}-
\widehat{\mathfrak{H}_{\mathfrak{R}R}^+}^{-1}\widehat{\mathfrak{H}_{\mathfrak{R}L}^+} \right)
\\
\widehat{\mathfrak{R}_{21}}=
\left[\widehat{\mathfrak{E}_{\mathfrak{R}R}^-}^{-1}\widehat{\mathfrak{E}_{\mathfrak{R}R}^+}-\widehat{\mathfrak{H}_{\mathfrak{R}R}^-}^{-1}\widehat{\mathfrak{H}_{\mathfrak{R}R}^+}
\right]^{-1}
\left( \widehat{\mathfrak{E}_{\mathfrak{R}R}^-}^{-1}\widehat{\mathfrak{E}_{\mathfrak{R}L}^-}-
\widehat{\mathfrak{H}_{\mathfrak{R}R}^-}^{-1}\widehat{\mathfrak{H}_{\mathfrak{R}L}^-} \right)
\\
\widehat{\mathfrak{R}_{22}}=
\left[\widehat{\mathfrak{E}_{\mathfrak{R}R}^-}^{-1}\widehat{\mathfrak{E}_{\mathfrak{R}R}^+}-\widehat{\mathfrak{H}_{\mathfrak{R}R}^-}^{-1}\widehat{\mathfrak{H}_{\mathfrak{R}R}^+}
\right]^{-1}
\left( \widehat{\mathfrak{E}_{\mathfrak{R}R}^-}^{-1}\widehat{\mathfrak{E}_{\mathfrak{R}L}^+}-
\widehat{\mathfrak{H}_{\mathfrak{R}R}^-}^{-1}\widehat{\mathfrak{H}_{\mathfrak{R}L}^+} \right)
\end{array}
\right.
\\
&\left\{
\begin{array}{l}
\widehat{\mathfrak{L}_{11}}=
\left[\widehat{\mathfrak{E}_{\mathfrak{L}L}^+}^{-1}\widehat{\mathfrak{E}_{\mathfrak{L}L}^-}-\widehat{\mathfrak{H}_{\mathfrak{L}L}^+}^{-1}\widehat{\mathfrak{H}_{\mathfrak{L}L}^-}
\right]^{-1}
\left( \widehat{\mathfrak{E}_{\mathfrak{L}L}^+}^{-1}\widehat{\mathfrak{E}_{\mathfrak{L}R}^-}-
\widehat{\mathfrak{H}_{\mathfrak{L}L}^+}^{-1}\widehat{\mathfrak{H}_{\mathfrak{L}R}^-} \right)
\\
\widehat{\mathfrak{L}_{12}}=
\left[\widehat{\mathfrak{E}_{\mathfrak{L}L}^+}^{-1}\widehat{\mathfrak{E}_{\mathfrak{L}L}^-}-\widehat{\mathfrak{H}_{\mathfrak{L}L}^+}^{-1}\widehat{\mathfrak{H}_{\mathfrak{L}L}^-}
\right]^{-1}
\left( \widehat{\mathfrak{E}_{\mathfrak{L}L}^+}^{-1}\widehat{\mathfrak{E}_{\mathfrak{L}R}^+}-
\widehat{\mathfrak{H}_{\mathfrak{L}L}^+}^{-1}\widehat{\mathfrak{H}_{\mathfrak{L}R}^+} \right)
\\
\widehat{\mathfrak{L}_{21}}=
\left[\widehat{\mathfrak{E}_{\mathfrak{L}L}^-}^{-1}\widehat{\mathfrak{E}_{\mathfrak{L}L}^+}- \widehat{\mathfrak{H}_{\mathfrak{L}L}^-}^{-1}\widehat{\mathfrak{H}_{\mathfrak{L}L}^+}
\right]^{-1}
\left( \widehat{\mathfrak{E}_{\mathfrak{L}L}^-}^{-1}\widehat{\mathfrak{E}_{\mathfrak{L}R}^-}-
\widehat{\mathfrak{H}_{\mathfrak{L}L}^-}^{-1}\widehat{\mathfrak{H}_{\mathfrak{L}R}^-} \right)
\\
\widehat{\mathfrak{L}_{22}}=
\left[\widehat{\mathfrak{E}_{\mathfrak{L}L}^-}^{-1}\widehat{\mathfrak{E}_{\mathfrak{L}L}^+}- \widehat{\mathfrak{H}_{\mathfrak{L}L}^-}^{-1}\widehat{\mathfrak{H}_{\mathfrak{L}L}^+}
\right]^{-1}
\left( \widehat{\mathfrak{E}_{\mathfrak{L}L}^-}^{-1}\widehat{\mathfrak{E}_{\mathfrak{L}R}^+}-
\widehat{\mathfrak{H}_{\mathfrak{L}L}^-}^{-1}\widehat{\mathfrak{H}_{\mathfrak{L}R}^+} \right)
\end{array}
\right.
\end{eqnarray}

Finally, we combine Eqs.(\ref{eqTMMgen}) and arrive to the generalization of the scattering-matrix technique that solves the problem of interlayer coupling: 
\begin{eqnarray}
\label{eqSMatGen}
\left\{
\begin{array}{llcllcl}
\vec{A}_L^- &= 
&\left(\hat{I}-\widehat{\mathfrak{L}_{12}}\widehat{\mathfrak{R}_{21}}\right)^{-1}
\widehat{\mathfrak{L}_{11}} &\vec{A}_R^-
&+
&\left(\hat{I}-\widehat{\mathfrak{L}_{12}}\widehat{\mathfrak{R}_{21}}\right)^{-1}
\widehat{\mathfrak{L}_{12}}\widehat{\mathfrak{R}_{22}} &\vec{A}_L^+
\\
\vec{A}_R^+ &= 
&\left(\hat{I}-\widehat{\mathfrak{R}_{21}}\widehat{\mathfrak{L}_{12}}\right)^{-1}
\widehat{\mathfrak{R}_{21}}\widehat{\mathfrak{L}_{11}} &\vec{A}_R^-
&+
&\left(\hat{I}-\widehat{\mathfrak{R}_{21}}\widehat{\mathfrak{L}_{12}}\right)^{-1}
\widehat{\mathfrak{R}_{22}} &\vec{A}_L^+
\end{array}
\right.
\end{eqnarray} 
\end{widetext}
with $\hat{I}$ being identity matrices. Eqs.(\ref{eqSMatGen}) represent the main result of this work.

Before illustrating the accuracy of the developed approach, we would like to underline its main advantages and disadvantages. The approach is ideal for calculation of light propagation in extended structures with relatively small number of multilayer segments; the increase of the number of segments results in additional memory use for each given segment and correspondingly, minimizes the advantages of wave-matching approach over finite-difference and finite-element schemes. 

The developed technique provides an effective mechanism for coupling the multilayers with high index contrast (high optical mode density difference) by implementing multilayer-dependent number of modes. However, our calculations show that careful design of spectrum of the open-waveguide modes is necessary when the coupling to and from the highly-confined modes is calculated. 

For guided modes, the approach allows straightforward calculation of inter-mode cross-talk by calculating the mode-specific pointing-flux and multiplying it by the amplitude of the given mode squared; similarly, the approach allow easy calculation of emission directionality, naturally separating the fields produced by guided modes from the fields of open-waveguide modes, and the fields of the incident waves from the fields of the scattered waves.

\section{Inter-guide coupling in plasmonic and metamaterial systems}
We now illustrate the accuracy of the presented approach on several examples of plasmonic and metamaterial systems. 

\subsection{Light emission and scattering by a waveguide}
We first consider light coupling to and from the waveguide. As an example, we use a $600$-nm-thick Si  waveguide surrounded by air, and calculate the coupling between this system and a homogeneous dielectric at $\lambda_0=1.5\mu m$; $\epsilon_{\rm Si}=12.12$\cite{palik}. 

\begin{figure}
\centerline{\includegraphics[width=8cm]{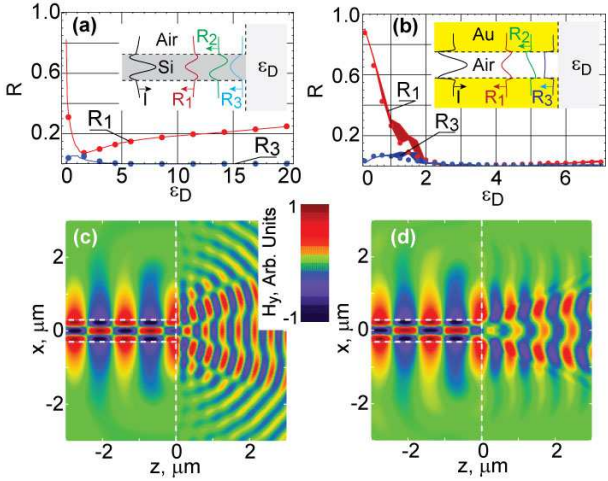}}
\vspace{10pt}
\caption{\label{figBulkScat} (color online) (a) light reflection in the planar air-Si-air waveguide coupled to a homogeneous dielectric; $\lambda_0=1.5\mu m$, waveguide thickness $d=0.6\mu m$; the geometry and profiles of the waveguide modes supported by the system are shown in the inset; excitation by $TM_2$ mode is assumed; the graph shows the comparison between the presented here technique (lines) and FEM simulations (dots); $R_2=0$ since the symmetry of $TM_1$ mode is different from that of $TM_0$ and $TM_2$ modes; (b) shows the reflection in plasmonic gap (Au-Air-Au) waveguide; $\lambda_0=d=0.6\mu m$; thickness of lines illustrates the convergence of the computations; panels (c) and (d) illustrate the field distributions in the system (a) with $\epsilon_D=12.12$ (c) and $\epsilon_D^{xx}=3.6 + 0.05 i, \epsilon_D^{yz}=-12.2 + 1.36 i$ (d)}
\end{figure}

To analyze the accuracy of our technique we assume that the system is excited by $TM_2$ mode with amplitude of $1$, and study the percentage of the reflected light into $TM_0, TM_1$ and $TM_2$ modes as a function of dielectric permittivity of the homogeneous dielectric [see inset in Fig.\ref{figBulkScat}(a)]. For comparison, we have calculated the same parameters with commercial finite-element-method (FEM) software\cite{comsol}. The perfect agreement between the results of our technique and FEM simulations is shown in Fig.\ref{figBulkScat}. As expected even for this relatively simple system, the scattering-matrix approach uses orders-of-magnitude less memory than FEM model. More importantly, scattering matrix formalism is easily extendable to the case of multiple guides or multiple interfaces. 

The field matching obtained with our technique for $\epsilon_D=12.12$ are illustrated in Fig.\ref{figBulkScat} (c). Panel (d) of the same figure illustrates the matching obtained in coupling between an air-Si-air guide and an anisotropic hyperbolic metamaterial with $\epsilon_D^{xx}=3.6 + 0.05 i, \epsilon_D^{yz}=-12.2 + 1.36 i$\cite{robynjosab,zhangNanowire,zayatsNanorod,chen,wegener,shalaevNATPHOT,pendrySuperlens,narimanovHyperlens,enghetaHyperlens,zhanghyperlens,smolyaninovhyperlens,sukoHG}.

\subsection{Light scattering in plasmonic systems}
To further analyze the accuracy and limitations of the developed field-matching technique we have calculated the scattering from the plasmonic analog of the Si-guide presented above: a $600$-nm-thick plasmonic gap waveguide operating at $\lambda_0=0.6\mu m$. We assume that cladding of this guide is composed from two gold strips with $\epsilon_{\rm Au}=-8.94 + 1.32 i$\cite{palik}. The agreement between the mode matching and FEM calculations of modal reflectivity in plasmonic gap guides is shown is Fig.\ref{figBulkScat}(b). 

\begin{figure}
\centerline{\includegraphics[width=8cm]{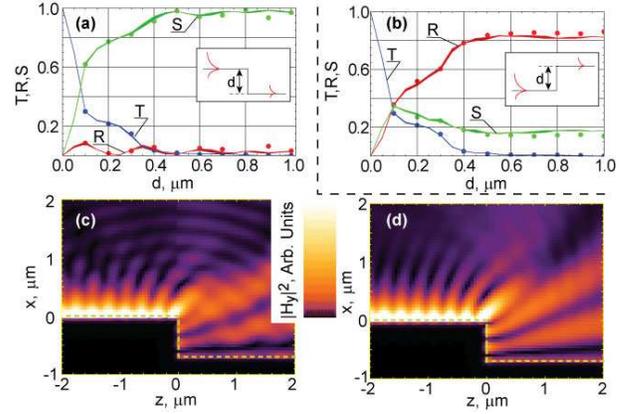}}
\vspace{10pt}
\caption{\label{figSPPscat}(color online) Scattering of the SPP propagating across Au-Air step; panels (a) and (b) show transmission, reflection, and scattering ($S=1-R-T$) of an SPP that is incident on the step; geometry of the structures is shown in insets; lines correspond to the formalism developed in this work (line thickness represents data variation due to changes in spectra of bulk modes), dots represent results from FEM simulations; panels (c) and (d) illustrate the field distributions obtained from scattering matrix (c) and FEM (d) simulations for $d=0.7\mu m$}
\end{figure}

We have further simulated the propagation of plasmonic modes across surface structure defects. In particular, we have used our approach to simulate the SPP propagation in ``plasmonic step'' geometry (Fig.\ref{figSPPscat}). As expected when the incoming SPP is travelling on the upper side of the plasmonic step, most of the incident energy is converted into free-space modes. In contrast, when the incident SPP is travelling on the lower side of the step, the majority of energy is converted into the reflected SPP wave.

Our numerical simulations demonstrate that at near-IR frequencies (when $|\epsilon_m|\gg\epsilon_d$), the scattered field can be successfully decomposed into ``top'' open waveguide modes, as suggested in Ref.\cite{oultonPRB}. However, in the proximity of SPP resonance ($\epsilon_m\simeq-\epsilon_d$), the inclusion of the ``bottom'' modes is necessary to adequately describe the optical properties of the system.

Note that the results of wave-matching simulations are almost identical to those obtained with FEM. Once again, we underline that wave-matching technique allows for calculation of light propagation in much larger systems than FEM system does.  

To assess the convergence of the our method, we have performed a set of simulations for each SPP structure described above, varying the configuration of spectra of top and bottom modes (for simplicity, equidistant $k_x$ spectra between $500$ and $3000$ modes were used). As expected, our simulations showed that it is necessary to design the spectrum of bulk modes to adequately resolve the SPP propagating at the $z=0$ interface. Interestingly, the ``averaged'' parameters (such as amplitudes of reflected guided modes) are much more sensitive to spectrum variations than the matching of the boundary conditions at $z=0$ interface, which is often considered to be an indication of accuracy of a numerical method. The typical inter-set variation of reflection, transmission, and scattering are illustrated in Figs.\ref{figBulkScat},\ref{figSPPscat}. The latter figure also shows the agreement between the field distribution obtained with our generalized scattering matrix formalism and with FEM. 

\section{Truly planar optics}
We now turn to the analysis of inter-mode coupling and out-of-mode scattering in planar optics. Planar optics in general (and SPP optics in particular) are fundamentally different from their free-space counterparts. Thus, when a plane wave is incident on the planar interface, the scattered field can be decomposed into one reflected plane wave, and one transmitted plane wave. In contrast to this behavior, when one guided mode is incident on the planar interface between two waveguides, it generates a continuum of open-waveguide modes in addition to the (sets of) reflected and transmitted guided waves. 

\begin{figure}
\centerline{\includegraphics[width=8cm]{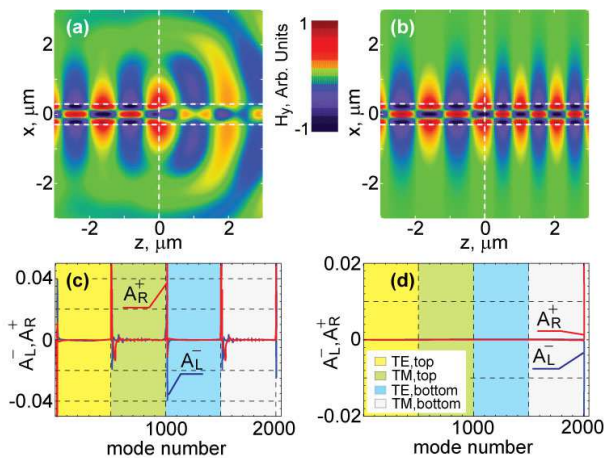}}
\vspace{10pt}
\caption{\label{fig2Doptics}(color online) (a,c) an interface between air-Si-air waveguide and isotropic air-$(\epsilon=6.06)$-air guide leads to substantial modal cross-talk, polarization mixing, and out-of-plane scattering; while the interface between air-Si-air waveguide and its anisotropic truly planar optics analog allows for ideal mode matching with light steering capabilities (b,d); guided modes in (c,d) correspond to mode number $>2000$; the system is excited by a $TM_2$ guided mode propagating at the angle $30^o$ to the $z=0$ interface; the amplitudes of modes in panels (c,d) are normalized to the amplitude of the incident mode}
\end{figure}

As seen from Figs.\ref{fig2Doptics},\ref{figLowScat} and from Ref.\cite{oultonPRB}, the typical interface between waveguide systems leads to scattering of about $20\%$ of incident radiation. Every attempt to change the effective index of the mode in planar structures is necessarily accompanied by modal cross-talk or by out-of-plane scattering of radiation. 


It is possible, however, to utilize anisotropic metamaterials to completely eliminate the cross-talk, and to map the familiar laws of 3D optics to optics of planar guides. The main idea of planar optics lies in the ability to guide light along the planar optical circuit with no out-of-plane scattering or modal cross-talk. In this section we assume that the layers on both sides of the interface are aligned with each other ($x_{L,j}=x_{R,j}$). 

In order to realize the efficient control over pulse propagation in the plane, the two layered structures must (i) have the same number of guided modes, and (ii) provide the ability to independently control the index of the mode [crucial for steering the light] and modal profile [crucial for optimizing the overlap integrals involved in $\widehat{\mathfrak{E}}$ and $\widehat{\mathfrak{H}}$ matrices]. 

As shown in Ref.\cite{elserprl} on the example of surface waves, these conditions can be satisfied when 
\begin{eqnarray}
\label{eq2Doptics}
\epsilon_{L_j}^{yz}&=\epsilon_{R_j}^{yz}
\\ \nonumber
n^2 \epsilon_{L_j}^{xx}&= \epsilon_{R_j}^{xx},
\end{eqnarray} 
with $n$ being the constant number that does not depend on layer number $j$. 

As can be explicitly verified, when Eq.(\ref{eq2Doptics}) is satisfied, all $\widehat{\mathfrak{E}}$ and $\widehat{\mathfrak{H}}$ matrices become diagonal. Thus, the inter-mode coupling is absent across the interface. The interface remains completely transparent to TE-polarized waves, while reflection and refraction of TM-polarized modes are controlled by the ratio of out-of-plane permittivities. The direction and amplitudes of the reflected and refracted modes are related to the direction and amplitude of the incident modes via the following Snell's law: 
\begin{equation}
 \label{snells}
\sin(\theta_i)= \sin(\theta_r)=n \sin(\theta_t)
\end{equation}
and Fresnel coefficients:
\begin{equation}
 \label{fresnel}
\frac{A_L^{(m)^-}}{A_L^{(m)^+}}=\frac{k_{L_z}^{(m)}-k_{R_z}^{(m)}}{k_{L_z}^{(m)}+k_{R_z}^{(m)}},
~~~
\frac{A_R^{(m)^+}}{A_L^{(m)^+}}=\frac{2k_{L_z}^{(m)}}{k_{L_z}^{(m)}+k_{R_z}^{(m)}}
\end{equation}
The above equations represent generalization of the formalism of Ref.\cite{elserprl} to multilayered guides. 

The concept of truly 2D optics is illustrated in Fig.\ref{fig2Doptics} on example of air-Si-air system coupled to metamaterials waveguide. As expected, reflection of a single mode in conventional planar system is accompanied not only by significant radiation scattering and modal cross-talk, but also by cross-polarization coupling. In contrast, for metamaterial guides the single incident mode excites one reflected wave, and one transmitted wave. 

\subsection{Low-scattering plasmonics}
An interesting extension of the truly planar optics is possible in plasmonic systems. From the fabricational standpoint, it is highly desirable that the plasmonic circuit is fabricated on top of common metallic substrate. 

Fabrication of extremely low-scattering plasmonic circuit is possible when using anisotropic dielectrics deposited on noble metals in the limit of visible and near-IR frequencies, where the permittivity of metal is much larger than the permittivity of dielectric ($|\epsilon_m| \gg\epsilon_d^{xx},\epsilon_d^{yz}$).

\begin{figure}
\centerline{\includegraphics[width=8cm]{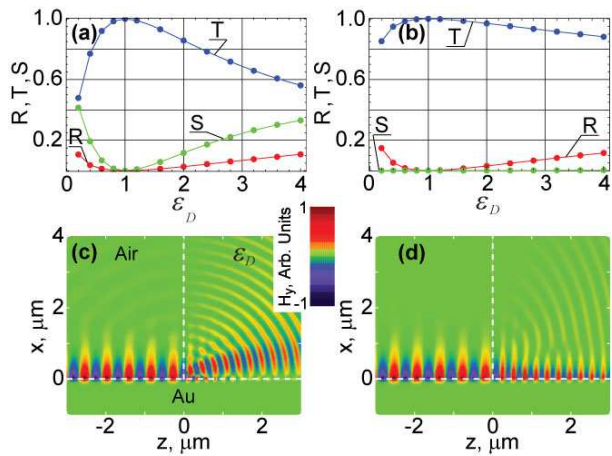}}
\vspace{10pt}
\caption{\label{figLowScat}(color online) Anisotropic coatings (b,d) can significantly reduce (and almost eliminate) the scattering losses in plasmonic circuits; the figure shows reflection, transmission, and scattering in conventional plasmonic circuit (a); field distribution for $\epsilon_D=4$ (c) clearly shows that only a fraction of the energy of incident SPP is transferred into SPP on the right-hand-side of the interface; on the other hand, the interface between anisotropic system with $\epsilon_{yz}=1, \epsilon_{xx}=4$ allows for substantial modulation of SPP index, and results in almost perfect SPP-SPP coupling across $z=0$ interface}
\end{figure}

Consider the situation where the in-plane ($yz$) component of permittivity of dielectric is kept constant across plasmonic system, and only out-of-plane ($xx$) component of permittivity of the superstrate is  varied. In such a system the propagation constant of the plasmonic mode, approximately given by\cite{podolskiycleo} 
\begin{eqnarray}
k_y^2+k_z^2 \simeq \epsilon^{xx}_d\left(1+\frac{\epsilon_d^{yz}}{|\epsilon_m|}\right)\frac{\omega^2}{c^2},
\end{eqnarray}
can be effectively controlled by changing the parameter $\epsilon_d^{xx}$. 

In the same limit, the exponential decay of the mode into the dielectric $\left[ E,H\propto \exp(-\kappa_d x), \kappa_d\simeq \omega/c \sqrt{\epsilon_d^{yz^2}/|\epsilon_m|}\right]$ does not depend on its propagation constant. The only source of out-of-plane scattering in such a structure is related to a weak dependence of the field profile in metal $[E,H\propto \exp(\kappa_m x)]$ on $\epsilon_d^{xx}$: $\kappa_m\simeq \omega/c\sqrt{|\epsilon_m|(1-\epsilon_d^{xx}/|\epsilon_m|)}$. 

As seen from Fig.\ref{figLowScat}, anisotropy of dielectric superstrate, achievable, for example, with electrooptic effect\cite{podolskiycleo}, provides orders-of-magnitude suppression of out-of-plane scattering in plasmonic systems with respect to isotropic counterparts.

\section{Conclusions}

To conclude, we have developed a reliable numerical technique for calculation of light propagation in planar guides and in arrays of planar guides. We have illustrated the developed formalism on examples of photonic, metamaterial, and plasmonic guides, and presented an approach to utilize anisotropic metamaterials for minimization and elimination of modal cross-talk in planar optical circuits. 

This work was sponsored by the ONR (Grant No. N00014-07-1-0457), NSF (Grant No. ECCS-0724763), and AFOSR (Grant No. FA9550-09-1-0029).




\end{document}